# Pricing and Revenue Sharing for Secondary Data Market

Hengky Susanto[1], Bhanu Kaushik[1], Benyuan Liu[1], Honggang Zhang[2] and ByungGuk Kim[1]

[1] Department of Computer Science, University of Massachusetts at Lowell.

[2] Engineering Deptement , University of Massachusetts at Bostom.

*{hsusanto, Bliu, Bkaushik, Kim}@cs.uml.edu, honggang.zhang@umb.edu*

**Abstract— Recent advances in technology enable public or commercial establishments (for e.g. malls, coffee shops, airports) and individual data plan subscribers to operate as Wi-Fi providers (WFP), offering Internet access. However, the model of a monthly flat service fee charged by ISP to establishment or individual WFPs offers very little incentives for ISP to provide additional bandwidth to WFP's customers. Thus, these users may experience insufficient bandwidth when there is high demand for bandwidth. In addition, ISP even discourages individual subscribers to provide Internet access through their smartphone because such practices may cause market saturation. To address this, in this paper we propose a dynamic pricing scheme and a revenue sharing mechanism that provide incentives for ISP to support establishment and individual Wi-Fi providers offering Internet access to users. Our revenue sharing model is based on Shapley value mechanism. Importantly, our proposed revenue sharing mechanism captures the power negotiation between ISP and Wi-Fi providers, and how shifts in the power balance between the two entities influence revenue division. Specifically, the model assures that the party who contributes more receives a higher portion of the revenue. In addition to that, the division of revenue eventually converges to a percentage value**.

## I. INTRODUCTION

To cater to the rising demand for Internet access, there is a growing number of public and commercial establishments offering Internet access, such as airports, coffee shops, libraries, etc. In addition to these, smartphones now have built-in Wi-Fi hotspot function that allows individual subscribers to provide Internet access. These establishments and subscribers can become secondary Wi-Fi providers and may offer their service for some monetary reward. However, there is very little incentive, and at the same time, there are challenges for Internet Service Providers (ISP) to support these alternative channels in their provision of Internet access to users. For example, allowing individual subscribers to provide Internet access, which we refer in this paper as *individual Wi-Fi provider* (I-WFP), means there will be more parties offering Internet access, making it more challenging for ISP to have an oversight of the market or control over pricing. From network management perspective, allowing I-WFP to provide such service may cause the market for Internet access to become saturated. Another example, *establishment Wi-Fi providers* (E-WFP) like malls or coffee shops commonly offer the Internet service free-of-charge or for a flat price. Both free-of-charge or flat fee arrangements offer little economic incentives for the E-WFP as well as ISP to offer additional bandwidthfQ to their users. Importantly, both E-WFP and I-WFP rely on ISP to provide the Internet access, i.e. ISP is ultimately responsible for the traffic generated by Wi-Fi users of E-WFP and I-WFP. In order to leverage on the emergence of E-WFPs and I-WFPs as secondary Wi-Fi providers, we will need to overcome the incentive barrier for ISP to enable such service and to provide Additional bandwidth. In this paper, we propose a pricing scheme and revenue sharing mechanism that provide incentives for ISP to support these secondary Wi-Fi providers and to offer additional bandwidth to their users. Our pricing model and revenue sharing mechanism collectively address concerns on network traffic load and price. Importantly, the revenue sharing mechanism assures economic incentives for both ISP and these secondary Wi-Fi providers.

We investigate the revenue sharing mechanism where secondary Wi-Fi providers share their revenue gained from providing the internet access to users with ISP. We build our revenue sharing model based on *Shapley value* mechanism [1,2] because of its capacity to divide the revenue "fairly" between parties involved. Our study shows that the cooperative scenario provides more incentives to ISP and secondary Wi-Fi providers to offer Additional bandwidth to users. That is, our cooperative revenue sharing model ensures that ISP's share of revenue increases as the secondary Wi-Fi providers generate higher income.

Following the revenue sharing model, we explore the pricing mechanism. Here, we argue that dynamic pricing strategy provides better economic incentives to both ISP and secondary Wi-Fi providers than a flat fee or free service. Our pricing model is built upon two principles: firstly, the pricing fluctuates according to the level of bandwidth demand. For ISP to set the minimum sale price to a secondary Wi-Fi provider, ISP must consider the total traffic load generated by these Wi-Fi providers as well as those generated by ISP's direct subscribers. Similarly, a secondary Wi-Fi provider computes the price to its users according to the level of demand, i.e. price rises as the level of demand increases. The second principle is, the final price charged to users of secondary Wi-Fi providers must be equal or higher than the minimum price set by ISP. This is because the minimum price set by ISP considers the overall bandwidth demand on the network, not just the demand from the users of secondary Wi-Fi providers. Hence, when the minimum price set by ISP is higher than the price set by secondary Wi-Fi providers, ISP's price prevails and users are charged with price set by ISP. These pricing frameworks are formulated into Network Utility Maximization (NUM) problem [3,4,5], which is resolved using subgradient based algorithm.

As secondary Wi-Fi providers, there is however a significant difference between establishment-WFP and individual-WFP: E-WFP pays a monthly fee for an unlimited amount of data usage, while I-WFP pays a monthly fee for a set quota of data. Therefore, the revenue sharing and pricing mechanism are tailored to incorporate this difference. For instance, the revenue sharing between ISP and E-WFP is designed to incentivize both entities to offer Additional bandwidth to E-WFP's users. On the other hand, revenue sharing between ISP and I-WFP is designed to provide incentive for ISP to allow I-WFPs to sell the unused bandwidth; at the same time, it discourages I-WFPs from achieving significant profitability from the transactions to prevent market saturation, as discussed in section II.

For reselling of bandwidth by E-WFP, our study shows that this pricing model provides incentive for both ISP and E- WFP to offer Additional bandwidth to their users. In addition, our revenue sharing model also discourages E-WFP's users from transmitting more data when ISP is experiencing congestion at ISP's end. Another important observation is that our revenue sharing model also captures the shift in bargaining power between ISP and E-WFP according to the amount of revenue generated. Specifically, the revenue sharing model apportions a larger share of the revenue to ISP when E-WFP is generating a lower revenue, and this is understandably so because ISP has to incur a minimum infrastructure overhead cost at all times. When E-WFP generates higher revenue from its users, our model attributes a higher portion of the revenue in recognition of its more significant contribution. However, the division of revenue for each E-WFP and ISP converges to a percentage value even when E-WFP generates most of the revenue, never to the disadvantage of the ISP.

As for I-WFP, our investigation emphasizes on a *cooperative* scenario where revenue is dynamically shared between ISP and the I-WFP, depending on the final sale price to users and traffic condition. This revenue sharing model is built based on *Shapley value* mechanism [1] because of its capacity to divide the revenue "fairly" between parties involved. Our study shows that such revenue sharing arrangement provides more incentives to ISP to support I-WFP and its users. Critically important is that this revenue sharing model ensures that ISP receives the majority portion of the revenue gained, regardless of the amount of the final sale price set by the I-WFP to prevent subscribers from seeking and achieving profitability (after deducting their monthly subscription fee) from selling their bandwidths. This is to avoid market saturation. Furthermore, our simulation shows that this model discourages I-WFP from selling their bandwidth during peak hours and provides the upper bound of the percentage I-WFP may receive.

Outcomes from our investigation, relating to both E-WFP and I-WFP, confirm that our revenue sharing model achieves "fair" compensation to all parties involved including ISP, as according to Shapley value properties

The rest of this paper is organized as follows. We begin with the background study in section II where we discuss related works, the impact on ISP, and a discussion on the different types of pricing and revenue sharing approaches. In section III, we present our proposed revenue sharing mechanism between ISP and secondary Wi-Fi providers, starting with the general model that is relevant to both E-WFP and I-WFP, followed by specific customizations according to the traits of E-WFP and I-WFP. In the followings section, we present the dynamic pricing mechanism using the same format: beginning with the general principles applicable to both E-WFP and I-WFP, followed by separate customization to meet the different conditions of E-WFP and I-WFP. The simulation results are presented in section V, followed by concluding remarks.

## II. Background

### A. Related Work

In [6], the authors propose a pricing strategy based on *online mechanism design* (OMD) to provide Wi-Fi service to Starbucks' customers. Their pricing strategy is designed on the basis of users dynamically arriving and leaving the coffee shop in a period of time; additionally they also consider that users make certain decisions based on particular outcomes as time progresses, such as a customer may decide to leave the shop after he/she has finished his/her drink, or to stay longer for more drinks. In addition, their pricing strategy also requires users to reveal their true valuation of the Internet access and their arrival time at the coffee shop. Our proposal, on the other hand, does not require users to reveal their valuation of the service to WFP, and price is determined according to network traffic, and is available after users start transmitting data. Our approach provides more flexibility and the price can be updated dynamically in real time setting. Furthermore, the role of ISP is not incorporated in the design proposed in [6]. Our model, on the other hand, allows ISP to influence WFP's pricing, especially when ISP is experiencing high traffic demand.

Revenue sharing between ISPs utilizing Shapley value has been studied in numerous literatures. For example, [7] has studied Shapley value to model ISPs' routing and interconnection decision. Similarly, [8] explores the design of profit sharing mechanism using Shapley value that allows the revenue to be decided "fairly" among participating ISPs. In [9], the authors examine the bilateral prices that can achieve the Shapley-value solution in ISP peering. These literatures primarily focus on revenue sharing between ISPs that are at par or almost equal in stature. In other words, multiple ISPs are partnering and collaborating at equal or almost equal level to deliver massive amount of data from content providers (like Google, CNN, ESPM, etc) to a very large pool of users. Our paper is mainly concerned with the revenue sharing between ISP and WFP, the latter being a customer of ISP, who relies on and pays ISP for Internet access. As WFP's network infrastructure is made up of only consumer level routers serving a much smaller pool of users, equal partnership between ISP and WFP is not possible in this setup.

Other than Shapley value based models, [10] proposes asymmetric Nash Bargaining Solution revenue sharing model between different types of ISPs. In this study, Stackelberg game, a *non-cooperative* game model, is considered in their pricing scheme, where a group of ISPs decide their prices according to the prices being offered in the market by other ISPs. In contrast, Shapley value based revenue sharing model is a *cooperative* based game theory [1]. In our discussion later we show that the cooperative model is more favorable for ISP, and yields a higher revenue for both ISP and WFP.

Finally, the authors of [11] propose a revenue sharing mechanism between *global* WFPs (*Skype*) and *local* WFPs (coffee shops, hotels, etc.) and the mechanism to incentivize local WFP to support Skype. In this setup, Skype users completely rely on WFP to provide the Additional bandwidth. Our model, in contrast, requires collaboration between WFP and ISP to provide Additional bandwidth. Furthermore, our proposed revenue sharing model divides the revenue according to the contribution of the participating parties.

### B. Impact on ISP

In order to design a suitable pricing and sharing mechanism, the impact of traffic from E-WFP and I-WFP on ISP must be considered. Users of these WFPs may increase the amount of data being injected into the network; especially during peak hours, the additional traffic may cause network to become even more congested, putting more pressures on ISP. Moreover, a higher traffic load may also negatively affect the connection quality of E-WFP's and I-WFP's users and degrade network performance. From a commercial perspective, while E-WFP and I-WFP may benefit financially from the fees charged to their users, ISP may have to incur higher costs to support the additional traffic without net additional financial benefits.

There is another concern relating particularly to I-WFP: network access through I-WFP becomes widely available at low prices, subscribers with low data usage may potentially cease their monthly subscription and choose to buy their network access from I-WFP as and when they need the connection. Higher demand for such ad hoc connection in turn may encourage some users to subscribe to data plan with the intention to resell bandwidth for profit. Over time, this may result in market saturation and loss of revenue for ISP. Given all these concerns and the potential ripple effects, the general opinion is that there is very little incentive for ISP to allow their subscribers to sell their unused bandwidth to reduce data plan underutilization.

Pertaining to the concerns and challenges raised, we propose a pricing and revenue sharing strategy that not only assure economic incentives for ISP, but also address network congestion. The pricing mechanism is made up of two parts: First, the pricing strategy between ISP and E-WFP/I-WFP where ISP determines the minimum price to resell the bandwidth at; simultaneously, E-WFP/I-WFP compute and decide their final sale price to the prospective users based on the level of their demand. Following that, we address the revenue sharing mechanism between ISP and E-WFP/I-WFP in both cooperative and non-cooperative scenarios.

### C. Free Service, Flat-rate, and Dynamic Pricing

To decide which pricing mechanism is appropriate to support E-WFP or I-WFP and provide Additional bandwidth, we first explore the tradeoffs between free service, flat-rate, and dynamic pricing. Free Internet access is commonly employed by E-WFPs to entice customers to boost their business sales. For example, a coffee shop offers free Wi-Fi to attract more customers to come and linger, with the objective of achieving higher sales per customer. E-WFP has very little incentive to provide Additional bandwidth because it does not gain obvious additional financial benefit from users enjoying good connection, especially when good connection has little direct impact on increasing the sale of their primary product/service. Similarly, flat rate pricing strategy for Internet access also provides little incentive for E-WFP to offer Additional bandwidth. For example, a hotel guest may pay a daily flat-rate ($20 per day) during his/her stay, and providing good connection does not increase the revenue because the guest still pays the same rate regardless of the quality of the connection. By the same token, there is little incentive for ISP to offer Additional bandwidth if E-WFP pays a flat-rate to ISP. As a result, without ISP's support for Additional bandwidth, E-WFP may not be able to offer Additional bandwidth even if it desires to. The same argument also applies to I-WFPs so that free service and flat rate are less desirable for offering Additional bandwidth.

Dynamic pricing is a usage-based pricing strategy where users pay according to the amount of bandwidth they use. This pricing model may provide more incentive to E-WFP and I-WFP to offer Additional bandwidth because users must pay more for a better connection quality, which leads to higher earnings for the providers. This pricing strategy can also be used to prevent congestion when there is more demand than available bandwidth by increasing the price to reduce traffic load. Therefore, dynamic pricing not only offers more opportunity for higher revenue, it also gives Wi-Fi providers better control over traffic. Similar arguments apply to ISP on the employment of dynamic pricing to support provision of Additional bandwidth to users of E-WFP and I-WFP.

### D. Cooperative Versus Non-Cooperative Strategies

In order to investigate what revenue sharing mechanism may provide better incentive for ISP to support E-WFP and I-WFP for Additional bandwidth, we explore both *cooperative* and *non-cooperative* strategies.

**Assumption 1**. Price charged $\lambda$ by Wi-Fi providers, both E-WFP and E-WFP, to users is at least the price charged $g$ by ISP, i.e. $\lambda \geq g$.

In the *non-cooperative* setting, ISP determines and charges Wi-Fi provider with the price $g$. Then, Wi-Fi provider sells the bandwidth to users at price $\lambda$ and pays ISP price $g$, where $g \leq \lambda$. Here, Wi-Fi provider gets to keep the difference between $\lambda$ and $g$. In all circumstances, ISP will not know Wi-Fi provider's final sale price $\lambda$ to users. This scenario may not be favorable to ISP especially when final sale price $\lambda$ is significantly higher than the price ISP charged to Wi-Fi provider $g$ because ISP does not get a proportionately higher share of the revenue. This non-cooperative strategy may also not be favorable to users because ISP does gain any additional financial benefit from allocating more resources for better connection. As a result, users may experience poor performance during peak time even after paying a high price to WFP. Thus, non-cooperative model is not supportive to user demand for Additional bandwidth.

In the *cooperative* setting, in addition to ISP deciding the minimum selling price $g$ to Wi-Fi provider and Wi-Fi provider setting the final sale price $\lambda$ to their users, ISP and WFP share the total final revenue received at price $\lambda$. In this model, ISP is informed of the final sale price $\lambda$ to users, and its portion of the revenue corresponds to the final price. By having visibility of the final sale price to users, ISP is able to monitor and assess the actual demand and value of the service. To fully develop this concept, we propose a revenue sharing mechanism based on Shapley value [6]. This mechanism is desirable because it exhibits several fairness properties that ensure revenue sharing is proportional to each party's contribution to the value of the revenue.

## III. Revenue Sharing

In this section, we address the mechanism of how the revenue gained from selling bandwidth to users is shared between Wi-Fi providers and ISP using revenue sharing mechanism based on Shapley value [1]. We first begin with the general principles that apply to both E-WFP and I-WFP, referred generically as WFP in this section.

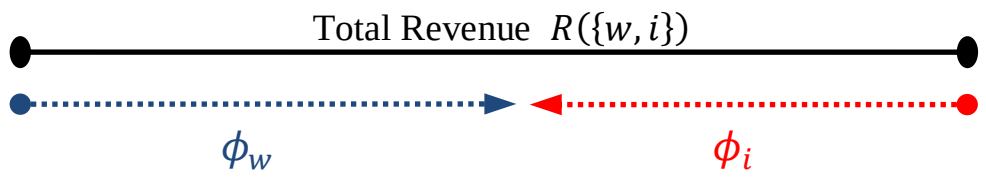

Fig. 1. Shapley value based revenue sharing.

Figure 1 generally illustrates how revenue $R$ is shared: $\phi_w$ apportioned to WFP $w$ and $\phi_i$ to ISP $i$, depending on their relative contribution.

### A. Desirable Sharing Properties

The design of the revenue sharing mechanism should satisfy these following properties. Let $R(.)$ be the revenue function and variable $\phi$ denotes a vector of *Shapley value*. Let $R(\{w,i\})$ denote the total revenue from providing network service by WFP $w$, where $w$ subscribes bandwidth from ISP $i$. The Shapley value has the following desirable properties [1,2]:

*Property 1 (efficiency)*: $\phi_w + \phi_i = R(\{w,i\})$.

The efficiency property requires the revenue shared to equal to the revenue from the service. In other words, the mechanism does not contribute or receive extra revenue.

*Property 2 (Symmetry)*: If $R(w \cup i) = R(i \cup w)$, then $\phi_w = \phi_i$.

The symmetry property requires that when ISP and WFP each contributes equally to the generation of the revenue, both should receive same portion of the revenue.

*Property 3 (Dummy player)*: If $w$ is a dummy WFP, $R(w \cup i) = R(i)$ and $\phi_w = 0$.

This property assures that when WFP $w$ is not contributing, then WFP $w$ receives zero share. Since WFP $w$ relies on ISP's infrastructure to sell bandwidth to users, ISP always has a contribution in supporting the network service, but not necessarily the WFP $w$.

*Property 4 (fairness)*: For any WFP $w$ and ISP $i$, the portion of revenue share is proportional to their respective contribution to the total revenue gained from the sale. This property addresses the fairness of revenue sharing between any pair of $\langle w, \mathrm{i}\rangle$.

*Property 5 (additivity)*: Lets separate a transaction into two parts, such that $T = T_1 + T_2$. For any two transaction $T_1$ and $T_2$, $\phi_i(N, T_1 + T_2) = \phi_i(N, T_1) + \phi_i(N, T_2)$, where $N = \{w, i\}$ and $(N, T_1 + T_2)$ is defined by $(T_1 + T_2)S = T_1(S) + T_2(S)$ for every coalition of $S$.

The property addresses the process of getting the Shapley value. The premise is that the outcome from transaction $(N, T_1 + T_2)$ should be equal to the addition of two different transactions of $(N, T_1)$ and $(N, T_2)$. To illustrate the idea, imagine WFP receive revenue according to transaction $(N, T_1)$ on the first day and $(N, T_2)$ on the second day. Assume that $T_1$ and $T_2$ are independent. Then, WFP's share from both days should be the summation of revenue shared from both transactions. In other words, this property guarantees that if the revenue of the service provided by WFP is additive, then the distributed revenue is the sum of the revenue generated for providing the service.

### B. Shapley Value Methodology

In this section, we will address the revenue sharing implementation between ISP and WFP using Shapley Value.

**Definition 1**. The *Shapley value* $\phi$ is defined by

$$\phi_j = \frac{1}{|N|!} \sum_{\pi \in \Pi} \Delta_i\big(R, Z(\pi, i)\big), \qquad \forall j \in N, \tag{1}$$

where $N = \{w, i\}$, $Z \subseteq$N, and

$$\Delta_j\big(R, Z(\pi, j)\big) = R(\, Z \cup \{j\} - R(Z)\,), \tag{2}$$

where $j \in N$. Remark: Given $(N, R)$, consider a permutation on $\pi$ on the set $N$. Members of set $N$ appear to "collect" their revenue according to the ordering $\pi$. For each member in $N$, let $Z_\pi^j$ be the set of members preceding member $i$, where $Z_\pi^j \subseteq N$. The marginal contribution of member $j$ according to $\pi$ is $\Delta_j\big(R, Z(\pi, j)\big) = R\left(Z_\pi^j \cup \{j\} - R\big(Z_\pi^j\big)\right)$. Here, the Shapley value can be interpreted as the expected marginal contribution $\Delta_j(R, Z)$, where $Z$ is preceding $j$ in an uniformly distributed random ordering. Since in this model $|N| = 2$, that is $N = \{w, i\}$, the Shapley value for $w$ and $i$ can be resolved by the following approach:

$$\phi_i = \frac{1}{2} R(\{i\}) + \frac{1}{2}\big(R(\{w, i\}) - R(\{w\})\big) \tag{3}$$

$$\phi_w = \frac{1}{2} R(\{w\}) + \frac{1}{2}\,\big(R(\{w, i\}) - R(\{i\})\big), \tag{4}$$

for $R(\{w, i\}) = \phi_i + \phi_w$. Here, ISP $i$ will keep $\phi_i$ of the revenue, while WFP $w$ will receive $\phi_w$ of the revenue, as illustrated in figure 1.

Total revenue $R(\{w, i\})$ earned at the WFP's end is defined as follows.

$$R(\{w, i\}) = \sum_{s \in w} x_s \lambda_s^f \; . \tag{5}$$

Here $x_s$ and $\lambda_s^f$ denote bandwidth usage and the final price charged by WFP $w$ to user $s$ respectively. Therefore,

$$R(\{w,i\}) = \sum_{s \in w} x_s\, \lambda_s^f \; = \; \phi_w + \phi_i\,,$$

where the term $s \in w$ denotes user $s$ utilizing the Internet service provided by WFP $w$. Additionally, in Shapley value methodology, $R(\{w\})$ and $R(\{i\})$ are the contributions of the Wi-Fi provider and ISP to revenue $R(\{w,i\})$ respectively, but they also can be interpreted as the revenue that they will gain if they do not collaborate. Here, $R(\{w\})$ is determined by solving the following equation

$$R(\{i\}) = \sum_{s \epsilon w} x_s\, g_s\,, \tag{6}$$

where $g_s$ denotes the price determined by ISP to provide service to user $s$. Revenue $R(\{i\})$ can be interpreted as the cost charged by ISP to WFP at price $g_s$ for providing service to user $s$. In (6), it is indicated that ISP charges different price to different user. However, since WFP relies on ISP to provide the network access, when $R(\{w\}) = 0$, then $R(\{i\}) = R(\{w,i\})$, which means ISP keeps the entire revenue of $R(\{w,i\})$. Thus,

$$R(\{i\}) = \begin{cases} \sum_{s \in w} x_s\, \lambda_s^f\,, & R(\{w\}) > 0 \\ R(\{w,i\}), & R(\{w\}) = 0. \end{cases} \tag{7}$$

The following two sections describe how $R(\{w\})$ is defined separately for E-WFP and I-WFP to address how each specifically contributes to total revenue $R(\{w\})$.

### *C. The Contribution of E-WFP*

E-WFP's contribution $R(\{w_E\})$ to $R(\{w_E,i\})$ is determined as follows.

$$R(\{w_E\}) = \frac{\sum_{s\,\epsilon\, w} \left(\lambda_s - g_s\right) x_s}{\max\left(log\left(g_w\right), \beta\right)}\,, \tag{8}$$

where

$$g_w = \sum_{s \epsilon w_E} g_s$$

and $\beta > 1$ is a positive constant to guarantee the denominator is greater than 1. To assure that a portion of the revenue is allocated to E-WFP, the denomination factor in (8) is concave and flattened as $g_w$ increases. Moreover, the gap between $\lambda_s$ and $g_s$ is considered in (8) to assure E-WFP receives more if $\lambda_s$ increases.

In our design of $R(\{w_E\})$, E-WFP's contribution is proportional to cost incurred by ISP to provide the access. A higher cost $g_w$ for access results in a lower E-WFP contribution to the total revenue. Next, we show that the property of revenue sharing mechanism.

***Theorem 1***. The revenue sharing mechanism assures that ISP's revenue portion at least covers the cost of providing service to E-WFP, i.e.

$$\phi_i \geq \sum_{s \epsilon w} x_s\, g_s.$$

*Proof.* By assumption that $g_s \leq \lambda_s^f$, that $g_s \leq \lambda_s^f$, $\lambda_s^f - g_s \geq 0$. Thus, by comparing $\lambda_s^f - g_s$ and $R(\{w\})$ in e.q. (8), we have the following equality.

$$\sum_{s\,\epsilon\, w} \left(\lambda_s^f - g_s\right) \geq \frac{\sum_{s\,\epsilon\, w} \left(\lambda_s^f - g_s\right)}{\max\left(log\left(g_w\right), \beta\right)}.$$

Observe that in the equality above, as the $g_w = \sum_{s \epsilon w} g_s$ increases, the right side of the equality decreases quicker than the left side. Next, this equality can be derived further as follows.

$$\sum_{s\,\epsilon\, w} \lambda_s^f - \sum_{s\,\epsilon\, w} g_s \geq \frac{\sum_{s\,\epsilon\, w} \left(\lambda_s^f - g_s\right)}{\max\left(log\left(g_w\right), \beta\right)},$$

Which is also

$$\sum_{s \epsilon w} g_s \;\leq \sum_{s \epsilon w} \lambda_s^f - \frac{\sum_{s \epsilon w} \left(\lambda_s^f - g_s\right)}{\max\left(log\left(g_w\right), \beta\right)}.$$

By considering bandwidth $x_s$ allocated for every user $s$ that receives service from WFP $w$, equality above also implies.

$$\sum_{s \epsilon w} (x_s\, g_s) \;\leq \sum_{s \epsilon w} \left(x_s\, \lambda_s^f\right) - \frac{\sum_{s \epsilon w} \left(\lambda_s^f - g_s\right) x_s}{\max\left(log\left(g_w\right), \beta\right)}. \tag{9}$$

By combining (3) with (5), (6), and (8), we have

$$\phi_i = \frac{1}{2} \sum_{s \epsilon w} (x_s\, g_s) + \frac{1}{2}\left( \sum_{s \epsilon w} \left(x_s\, \lambda_s^f\right) - \frac{\sum_{s \epsilon w} \left(\lambda_s^f - g_s\right) x_s}{2\, \max\left(log\left(g_w\right), \beta\right)} \right). \tag{10}$$

Next, we substitute (9) to (10), such that

$$\phi_i \geq \frac{1}{2} \sum_{s \epsilon w} x_s\, g_s + \frac{1}{2} \sum_{s \epsilon w} x_s\, g_s = \sum_{s \epsilon w} x_s\, g_s.$$

Thus, $\phi_i \geq \sum_{s \epsilon w} \left(x_s\, g_s\right)$, which is the revenue shared apportion to ISP covers the minimum cost. ■

Additionally, by definition, E-WFP's minimum share of revenue is described by e.q. (8).

### *D. The Contribution of I-WFP*

Next, our proposal determines I-WFP's $R(\{w_I\})$ is designed to only allow users to offset some of the data plan subscription fee they pay each month, and not for profitability. The objective of this approach to reselling unused bandwidth is to somewhat alleviate the sentiment of money wasted because of underutilized data plan at the end of each month. Furthermore, $R(\{w_I\})$ is subjected to

$$R(\{w_I\}) \;\leq R(\{w_I, i\}) - R(\{i\}), \tag{11}$$

for $R(\{i\}) > 0$. Condition (11) is to assure ISP's share is at least $R(\{i\})$. Since the magnitude of $R(\{i\})$ influences how much revenue an I-WFP $w_I$ receives, $R(\{w_I\})$ is capped according to the law of *diminishing returns*, where the contribution to the total revenue diminishes as $R(\{\dot{s}\})$ grows. Thus, $R(\{w_I\})$ is concave and is defined as follows.

$$R(\{w_I\}) = \omega \log\left(\alpha\left(R(\{w_I, i\}) - R(\{i\})\right)\right), \quad (12)$$

where $\alpha$ is a positive constant variable decided by ISP, positive weight function $\omega = \frac{x_{unusedBW}}{x_{dp}}$. Here, $x_{unusedBW}$ denotes the unused bandwidth at time $t$ and $x_{dp}$ is the total of bandwidth included in the monthly data plan ($x_{dp}$ is replenished at the beginning of the billing cycle). Weight $\omega$ is an indication of the usage level, i.e. a user with higher $\omega_t$ implies a lower data usage and vice versa. Furthermore, $\omega$ is also used to limit reselling of bandwidth to low data usage subscribers in order to help them reduce the excess of data plan underutilization. Also, in order to prevent I-WFP from making profit after deducting their subscription fee at the end of the billing cycle, the total amount of revenue received by I-WFP from reselling transactions does not exceed their subscription fee. Thus,

$$R(\{w_I\}) = \begin{cases} 0, & h(w_I) \le \sum_{k=0}^{K} \phi_w^k, \\ eq(12), & Otherwise, \end{cases}$$

where $K$ is the number of reselling transactions per month cycle and $h(w_I)$ denotes the monthly fee paid by I-WFP $w_I$.

**Theorem 2**: I-WFP $w$ share of revenue decreases as $x_w \to x_{DP(w)}$.

*Proof*. We have $x_s \in x_w$ , for $\forall s \in w_I$, that is the amount of bandwidth used by I-WFP $w_I$ including the bandwidth he/she has sold. Without losing the originality, let $\lambda_s$ and $g_s$ be the price that I-WFP charges his/her client $s$ and the minimum price for $s$ determined by ISP respectively.

$$\frac{\partial R(\{W\})}{\partial x_w} = -\frac{1}{2x_{DP(w)}} \log\left(x_w \sum_{s \in w}\left(\lambda_s^f - g_s\right)\right) + \frac{x_{DP(w)} - x_w}{2\, x_{DP(w)}\, x_w \sum_{s \in w}\left(\lambda_s^f - g_s\right)}.$$

Notice that $\frac{x_{DP(w)} - x_w}{2\, x_{DP(w)}\, x_w \sum_{s \in w}\left(\lambda_s^f - g_s\right)} \to 0$, as $x_w \to x_{DP(w)}$ and $-\frac{1}{2x_{DP(w)}} \log\left(x_w \sum_{s \in w}\left(\lambda_s^f - g_s\right)\right) \le 0$. Thus, $\frac{\partial R(\{w\})}{\partial x_w}$ decreases, as $x_w \to x_{DP(w)}$, which also implies that $R(\{w\})$ decreases when $x_w \to x_{DP(w)}$. Thus, $\phi_w$ in (14) decreases as $x_w \to x_{DP(w)}$, which also satisfies property 1. That is, if $R(\{w\})$ decreases, then, $\phi_w$ also decreases. Next, consider a special case that $R(\{w\}) = 0$ when $x_w = x_{DP(w)}$. Constraint (16) satisfies property 4, such that $\phi_w = 0$, which means ISP keeps the entire revenue share. Thus, $\phi_w$ decreases as bandwidth usage $x_w$ increases.■

Proposition 3 confirms that subscribers with lower data usage will receive a higher share of revenue from the reselling transactions than subscribers with higher data usage. In addition, eq. (17) assures that the portion of reselling revenue gained by subscribers diminishes as the amount of bandwidth used/sold increases. Thus, subscribers with higher data usage have less incentive to sell their unused bandwidth.

It is natural for E-WFP or I-WFP to want to maximize his/her Shapley value in order to maximize the earning. However, it is very difficult to determine their maximum earning because maximizing Shapley Value is coNP-hard [2]. The solution to Shapley Value maximization problem is discussed in [2].

## IV. Pricing Mechanism

In this section, we address ISP's pricing mechanism that applies uniformly to both E-WFP and I-WFP, and the Wi-Fi provider's pricing to users, regardless if they are users of E-WFP or I-WFP. Therefore, in this section, both E-WFP and I-WFP are collapsed and referred to as WFP. Figure 2 below illustrates the overview of the transaction: after an user $s$ makes his/her the request for connection, ISP presents WFP with the *minimum price* $g_s$. At the same time, WFP computes the *WFP's price* $\lambda_s$, and determine the *final sale price* $\lambda_s^f$, as formulated

$$\lambda_s^f = max(\lambda_s, g_s + \rho), \quad (13)$$

where $\rho$ denotes a constant minimum profit decided by WFP, for $\rho \ge 0$. Then, WFP presents price $\lambda_s$ to user $s$ and user $s$ pays the WFP at price $\lambda_s$.

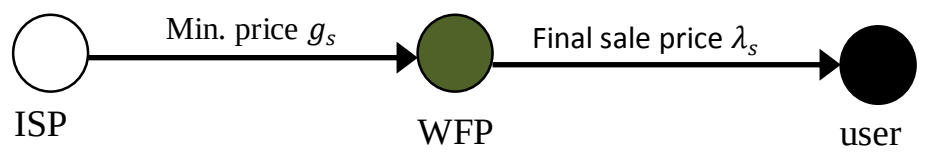

Fig. 2. Pricing mechanism.

Moreover, the overview of the pricing formulation is that ISP decides the minimum sale price $g_s$ to support user $s$ and at the same time WFP also decides the its own price $\lambda_s$ to user $s$. Then, user $s$ pays the service at price $\lambda_s^f$ (the maximum between the two prices according to e.q (10)), as illustrated in figure 3.

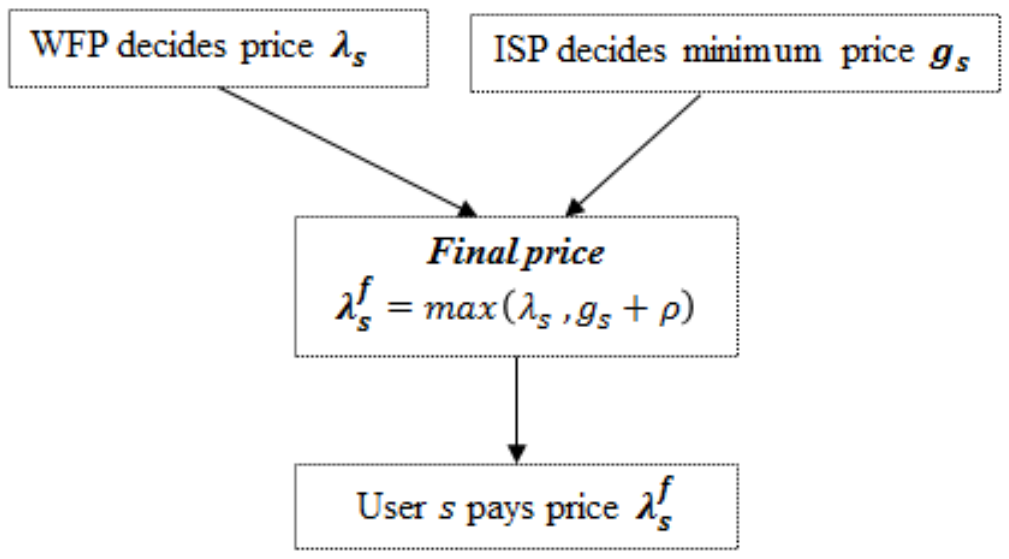

Fig. 3. Pricing Formulation.

The pricing mechanism also considers multiple users at any point of time. We begin by first addressing WFP's price to user.

### *A. User Utility Function*

Let $S_w$ denotes a set of user $s$ using the Internet, for $s \in S_w$. The objective of user $s$ is to solve

$$max\, U\left(x_s, \lambda_s^f\right), \quad \text{for } x_s, \lambda_s^f \ge 0,$$

where $x_s$ denotes the amount of data usage by user $s$ and $\lambda_s^t$ denotes the price to be paid by user $s$ for Internet access at time $t$. The price is dynamically determined according to the level of demand for network service. The utility function of the user is defined as follows.

$$U\left(x_s, \lambda_s^f\right) = U_{bw}(x_s) + U_{cost}\left(x_s, \lambda_s^f\right),$$

where $U_{bw}(x_s)$ and $U_{cost}\left(x_s, \lambda_s^f\right)$ denote user $s$ utility with bandwidth consumption $x_s$ and service cost, respectively. Considering the WFP is operating at frequency band $B_s$, the utility function of bandwidth usage is defined as follows

$$U_{bw}(x_s) = W_s \log\left(x_s\left(1+\frac{P_s\,|c_s|^2}{\partial_s^2\,B_s}\right)\right),$$

where $P_s$ is the transmission power of user $s$ mobile device, $c_s$ is the channel gained from WFP $w$ to user $s$, and $\partial_s^2$ is the Gaussian noise variance for the channel between $w$ and $s$ [14]. In other words, $U_{bw}(x_s)$ is influenced by the channel quality and the amount of bandwidth. Additionally, $U_{bw}(x_s)$ follows the law of diminishing returns. This is because more bandwidth does not always mean higher satisfaction and SNR measurement for wireless is concave [14].

Utility function $U_{cost}(x_s, \lambda_s^f)$ represents user satisfaction for monetary surplus when the cost paid for Internet access is less than the budget, which is defined as follows.

$$U_{cost}(x_s\,, \lambda_s^f) = 1 - \frac{x_s\,\lambda_s^f}{m_s},$$

where $m_s$ denotes the budget that user $s$ is willing to spend for bandwidth $x_s$. Note that $x_s\,\lambda_s^t$ can be interpreted as the price that user $s$ must pay for the service. Thus, ideally, user's budget matches the price that he/she must pay for the service, such that $\frac{x_s \lambda_s^f}{m_s} = 1$ and hence $U_{cost}(x_s\,, \lambda_s^f) = 0$. Therefore, given price $\lambda_s^f$, user $s$ utilizes $m_s$ to influence the amount of bandwidth $x_s$ allocated to him/her.

*B. Pricing Mechanism of WFP*

The objective of WFP $w$ is to maximize its own revenue without exceeding its monthly bandwidth capacity. The maximization problem is expressed as follows.

$$max \sum_{s\,\epsilon\,w} x_s\,\lambda_s\,, \tag{14}$$

$$s.t. \sum_{s\,\epsilon\,w} x_s \le C_w\,,$$

$$over\; x_s \ge 0, \;\; \forall s \in w,$$

where capacity $C_w$ is amount of WFP's bandwidth capacity. Considering the respective objectives of user and WFP, the problem can be formulated into network utility maximization (NUM) [2].

$$max \sum_{s\,\epsilon\,w} U(x_s, \lambda_s) \tag{15}$$

$$s.t. \sum_{s\,\epsilon\,w} x_s \le C_w\,,$$

$$\text{over}\; x_s \ge 0, \;\; \forall s \in w,$$

The Lagrangian optimization problem is formulated as

$$L(\bar{x}_s, \bar{\lambda}_s) = \sum\nolimits_{s\epsilon w} U(x_s, \lambda_s) - \sum_{s\epsilon w} x_s\,\lambda_s + \sum_{s\epsilon w} \lambda_s\,C_w,$$

where $L(.)$ is the Lagrangian form and $\lambda_s$ is known as the Lagrangian multiplier, which is often interpreted as the link price, and $\bar{x}_s$ is a vector of $x_s$, for $\forall s\;\epsilon\;w$, and $\bar{\lambda}_s$ is a vector of $\lambda_s$. The common solution to NUM problem is the subgradient based method [3]. Typically, the dual problem $D$ to the primal problem of (12) is constructed as follows

$$\min D(\bar{\lambda}_s)\,, \;\; s.t\;\; \bar{\lambda}_s \ge 0,$$

where the dual function

$$D(\bar{\lambda}_s) = \max_{0\le\bar{x}_s\le x^{max}} L(\bar{x}_s, \bar{\lambda}_s).$$

To solve $D(\bar{\lambda}_s)$, first user $s$ maximizes over $x_s$ given $\lambda_s$. That is

$$x_s = \underset{0\,\le\,x_s\,\le\,x_s^{max}}{\arg\max} \left(U(x_s\,, \lambda_s)\right). \tag{16}$$

However , since e.q. (13) assures the minimum price charged to users. Thus, e.q. (16) can be expressed as follows.

$$x_s = \underset{0\,\le\,x_s\,\le\,x_s^{max}}{\arg\max} \left(U(x_s\,, \lambda_s^f)\right)$$

Next, $L(\bar{x}_s, \bar{\lambda}_s)$ is minimized with subgradient projection method in an iterative solution given by

$$\lambda_s^{t+1} = \left[\lambda_s^t - \sigma^t\left(C_s - \sum_{s\epsilon w} x_s\right)\right]^+, \tag{17}$$

where $C_s - \sum_{s\epsilon S} x_s$ is a subgradient of $D(\lambda_s)$ and $\sigma^t$ denotes the step size to control the tradeoff between a convergence guarantee and the convergence speed, such that

$$\sigma^t \to 0, as\; t \to \infty\; and \sum_{t=1}^{\infty} \sigma^t = \infty. \tag{18}$$

Next, after solving $\lambda_s^{t+1}$, then we solve for $\lambda_f^{t+1}$ with (13). Notice that in (13), $g_s^{t+1} + \rho$ serves the minimum price charged to user $s$, such that $\lambda_s^{t+1} \ge g_s^{t+1} + \rho$. Generally, the subgradient based solution relies on feedback loop mechanism. That is, the user determines the transmission rate according to the price set by WFP by solving (16) and the price is adjusted according to the traffic load by solving (17). It is repeated until it converges to an optimal solution. Price $\lambda_s$ is also an indication of the demand for service. However, before determining the final sale price, WFP must consider the minimum price charged by ISP as described in (13). It is because WFP depends on ISP's infrastructure to provide the service. The discussion on the minimum price is addressed in the next section.

**Proposition 1:** If the step size $\sigma$ in (18) satisfies (17), then the subgradient based algorithm converges to the optimal solution of problem (15). [13] ■

*C. Minimum Price Set by ISP*

In this section, we address the pricing mechanism for ISP to determine the minimum price for users to get Internet access from either E-WFP or I-WFP. Consider a network managed by ISP with a set of links $L$, and a set of link capacities $C$ over the links in $L$. Given a utility function $U_s(x_s, \lambda_s)$ of data user $s$ with bandwidth usage of $x_s$ and traffic generated by users in $S$, the maximization problem can be formulated as follows.

$$max \sum\nolimits_{s\in S} U(x_s) + \sum_{\acute{s}\in\acute{S}} U(x_{\acute{s}}), \tag{19}$$

$$s.t. \quad \sum_{s\epsilon w} x_s \le C_w \,, \tag{19.a}$$

$$\sum_{s\in l} x_s + \sum_{\dot{s}\,\in\, l} x_{\dot{s}} \;\le C_l, \tag{19.b}$$

$$x'_{\dot{s}l} \;= x_{\dot{s},l} \,, \tag{19.c}$$

$$\text{over } x_{\dot{s}}, x_s \ge 0,$$

where $S$ is a set of users that get access from WFP, $\dot{S}$ denotes a set of ISP subscribers (users who get access directly from ISP), $\dot{s} \in l$ and $s \in l$ denotes WFP user $s$ and ISP user $\dot{s}$ who are transmitting data through link $l$. Conditions (19.a) and (19.b) guarantee the link is *feasible*, i.e. the total traffic is less than or equal to the link capacity. Condition (19.c) guarantees that the flows from the existing subscribers are not affected by traffic generated by WFPs' users. However, the constraint (8.c) makes problem (19) difficult to solve, or it may not have a solution. To resolve this, problem (19) can be simplified by relaxing the constraint (19.c) and substituting it with a new constraint $U_s(x_s) \ge 0$, for $s \in S$. This way, it allows ISP to be more flexible with the bandwidth allocation. However, problem (19) becomes a non-convex problem, which is an NP-Hard problem [12]. It is because users with real-time traffic is usually modeled using a sigmoidal function, which is non-concave. Notice that with the new constraint, the connection quality of the subscribers may be compromised when network experiences a bottleneck.

Hence, to resolve problem (19), we consider an equivalent problem. Since the objective on constraint (19.c) is to assure that Additional bandwidth of ISP subscribers are not affected by traffic generated by WFP users, $\sum_{\dot{s}\,\epsilon\, l} x_{\dot{s}}$ amount of bandwidth can be reserved for ISP subscribers. Thus, problem (19) can be simplified by combining constraints (19.a), (19,b), and (19.c). Hence, we consider the equivalent problem.

$$\max \sum\nolimits_{s\in S} U(x_s) \tag{20}$$

$$s.t. \sum_{s\epsilon l} x_s \;\le C_l - \sum_{\dot{s}\epsilon l} x_{\dot{s}},$$

$$over\ x_s \ge 0,$$

The constraint in problem (20) prevents the connection quality of ISP's subscribers from being compromised by traffic from WFP's buyers. Therefore, packets that belong to subscribers receive a higher priority than those of WFP's users during occurrence of network congestion. Problem (20) is similar to problem (15), which also can be resolved using sub-gradient based algorithm. To solve problem (20), users solve eq. (16) and ISP determines the minimum price to sell on each link $l$ by solving

$$g_l^{t+1} = \left[ g_l^t - \sigma^t \left( \left( C_l - \sum_{s\epsilon S} x_s \right) - \sum_{s\epsilon l}\sum_{s\epsilon w} x_{\bar{s}} \right) \right]^+ . \tag{21}$$

Here, $s \in l$ denotes user $s$ who is transmitting data through link $l$. The total *minimum price* to sell to user $s$ is $g_s = \sum_{s\epsilon l} g_l^t$, for $\forall s, s\epsilon S_w$.

## V. Simulation and Discussion

In this section, we analyze the behavior of revenue sharing between ISP and WFP. More specifically, it is to investigate how the difference between the final sale price paid by user and the minimum price set by ISP influences revenue sharing, and whether the outcome is favorable to ISP and/or WFP. The analysis is conducted in two separate case studies: revenue sharing between ISP and E-WFP, and between ISP and I-WFP to highlight the differences between E-WFP and I-WFP.

### *A. Revenue sharing between ISP and E-WFP*

In the simulation setup, we have a E-WFP subscribing Internet access from an ISP to provide Wi-Fi to users. In this setup, the E-WFP provides to users bandwidth of 10 MB/sec and the initial minimum price charged by ISP is 10 units currency and the total minimum profit desired by E-WFP is 5 units currency. Thus, the initial price charged to users is 15 units currency. We apply this set up to two scenarios we are investigating: firstly, when congestion occurs at the ISP's network, and secondly, when there is a high demand at the WSP's end but low traffic load in ISP's network. In each scenario, either ISP's or E-WFP's price is raised incrementally by one unit up to 300 increments.

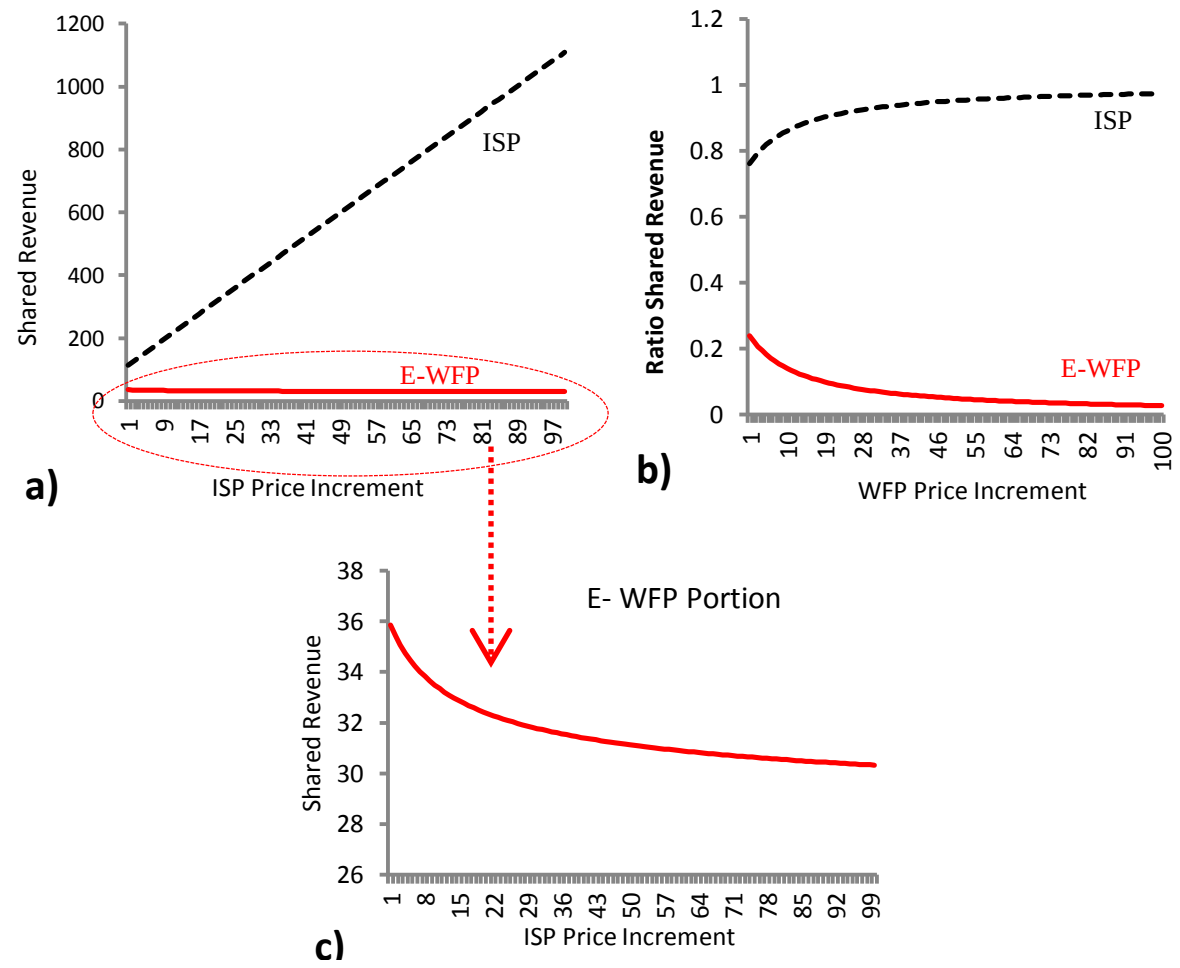

Fig. 4. Revenue sharing when ISP increases the price.

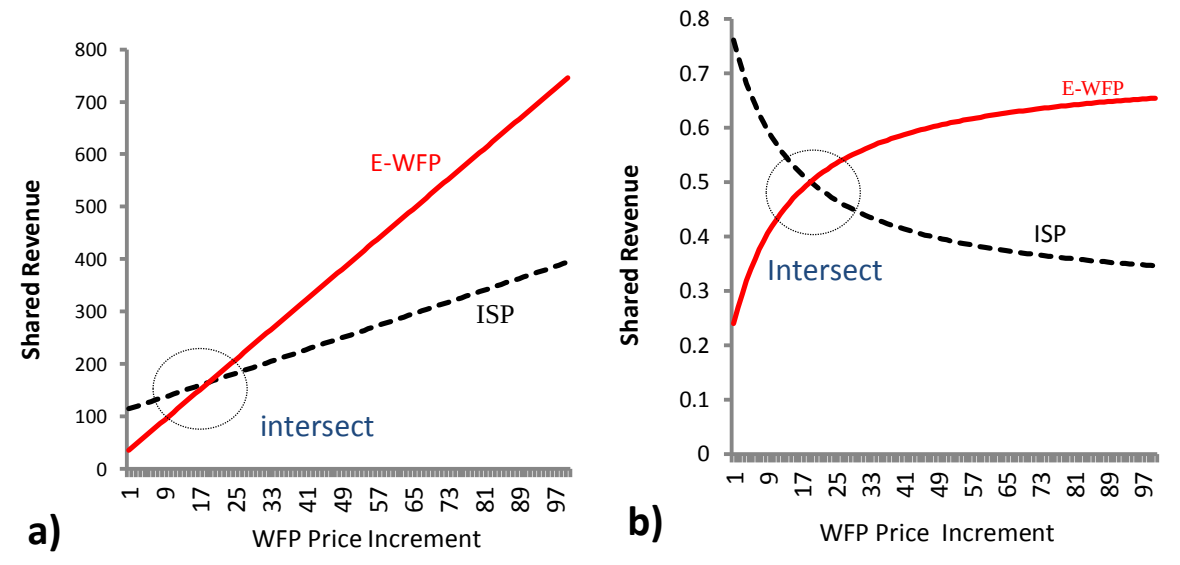

Fig. 5. Revenue sharing when E-WFP increases the price.

**First scenario**. During ISP's peak hours, ISP increases the price to reduce the amount of demand. In this scenario, there are ten users getting Internet access from a Wi-Fi provider, where each user receives equal amount of bandwidth allocation. Figure 4(a) and 4(b) illustrate that our proposed revenue sharing mechanism favors ISP during its peak hours, while at the same time ensuring that E-WFP receives some portion of the revenue. In other words, the majority of the revenue is allocated to ISP during ISP's peak hours. The y-axis of figure 4(a) depicts the revenue portion allocated to ISP and E-WFP in unit currency and the y-axis in figure 3(b) is the percentage of revenue apportioned to E-WFP and ISP, totaling to 100%. Therefore, there is little incentive for E-WFP to provide Additional bandwidth during peak hours. However, as figure 4(c) demonstrates, regardless of how much ISP charges, E-WFP always receives some portion of the revenue as E-WFP's portion converges to a value even when ISP's price continues to grow. In essence, the outcome from this scenario implies that during peak hours ISP receives most of the share of the revenue, regardless how much E-WFP charges its users.

**Second scenario**. E-WFP receives a high level of demand from users, causing E-WFP to increase its price to users. However, there is low traffic at ISP's end, where the minimum price decided by ISP is much lower than E-WFP's price to users. In this second scenario, there are initially ten users getting Internet access but eventually the number of users is increased by five users in each iteration. Figures 5(a) and 5(b) illustrate the division of revenue between E-WFP and ISP as E-WFP's price is incrementally raised by 1 unit currency up to 300 increments. Figure 5(a) depicts the division of revenue in unit currency and figure 5(b) in percentage value.

Figure 5(a) shows that in this scenario both ISP and E-WFP receive relatively higher revenue due to the high demand from E-WFP's users. The higher revenue received by ISP should provide an incentive for ISP to provide Additional bandwidth. From the first to the 22$^{nd}$ price increment, as shown in figure 5(a), ISP receives a higher share of the revenue than E-WFP because up to this point the revenue is still relatively low. This can be justified because ISP provides and manages the infrastructure, a higher portion of the revenue is allocated to ISP to cover the cost of providing the access and managing the traffic from E-WFP.

At 22$^{nd}$ price increment, the two lines intersect (figure 5). The intersection can be interpreted as when the bargaining power is balanced and revenue is equally shared between ISP and E-WFP. However, as E-WFP continues to increase price and generates higher revenue, and traffic at ISP's end remains low, the bargaining power progressively shifts toward E-WFP because of its higher "contribution" to the transaction. E-WFP's portion of revenue eventually converges to a region of 70% of the total revenue (figure 4(b)), and ISP's percentage share converges to 30%. Generally, the convergence to a specific value shows asymptotically that there is a predictable region of revenue sharing. Importantly, this convergence also provides the upper celling of the portion allocated to E-WFP, i.e. 70%, and the bottom limit of the portion allocated to ISP, i.e. 30%. The existence of these upper and lower bounds for revenue sharing can become the basis for both WFP and ISP to evaluate and negotiate the risk and gain in such trade agreement for mutual benefits.

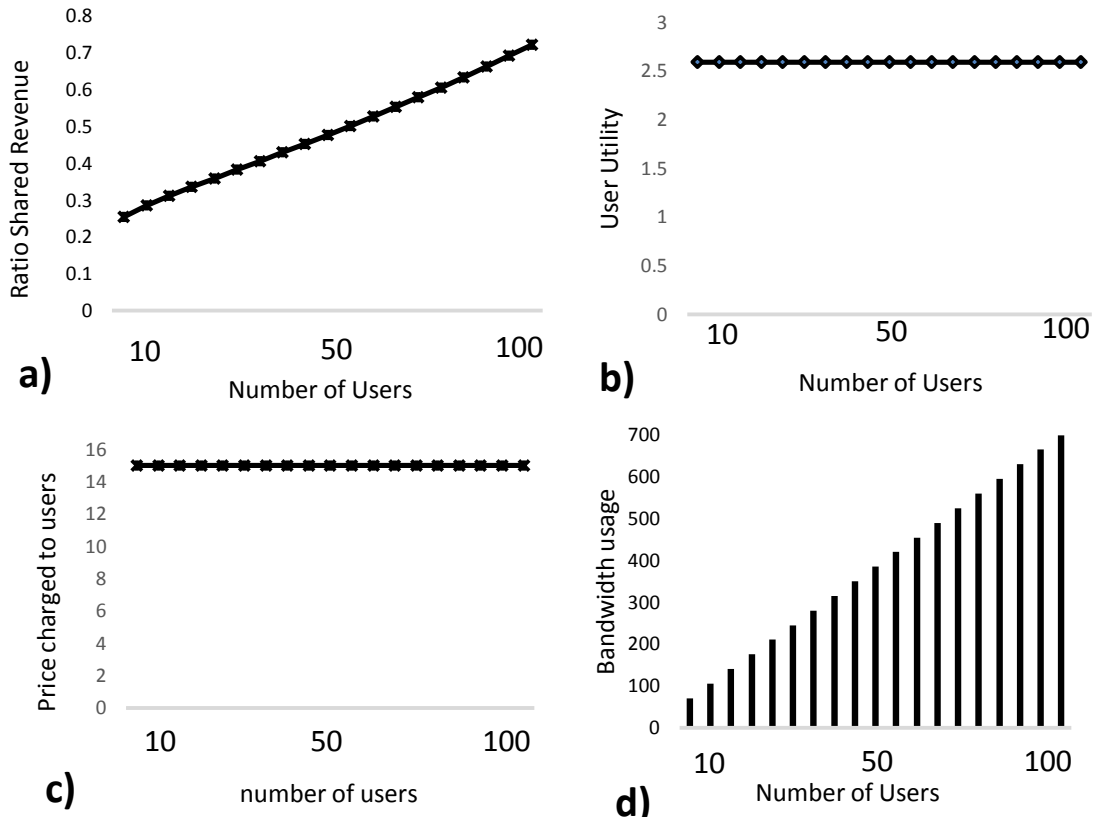

Fig. 6. Revenue sharing when bandwidth demand is *low* with up to 100 users. (a) Ratio shared revenue of WFP over ISP. (b) Average user utility. (c) The price that users pay for the service. (d) Amount of bandwidth sold to users.

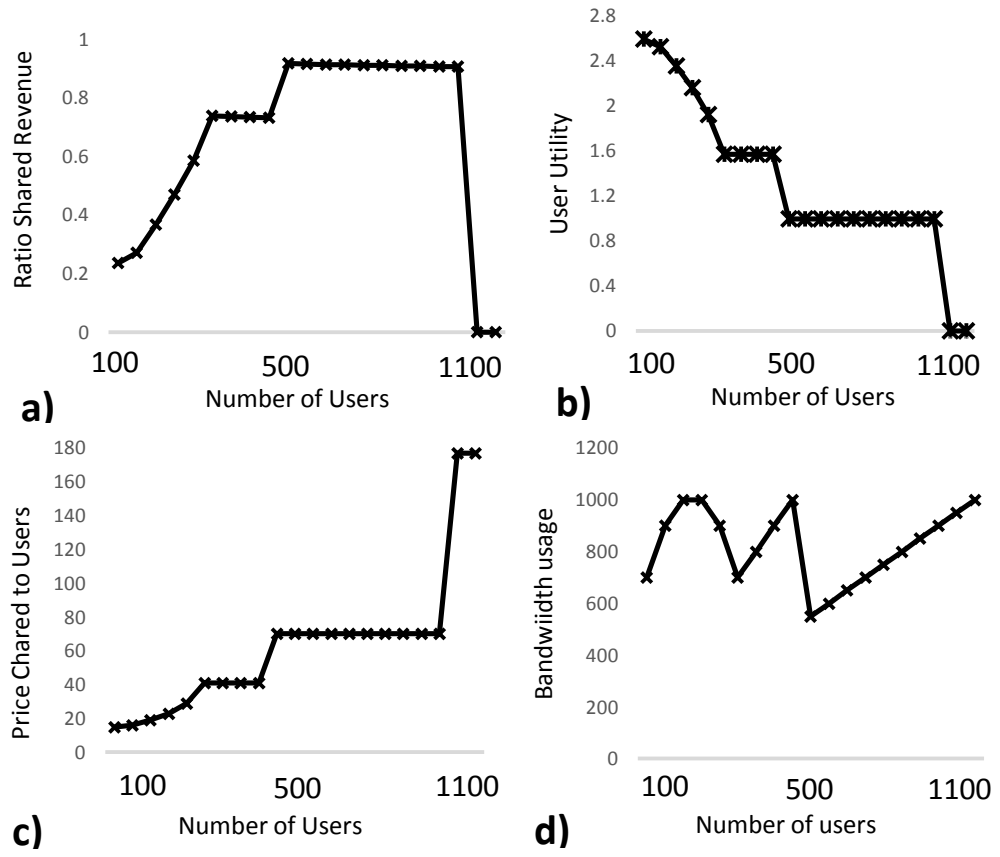

Fig. 7. Revenue sharing when bandwidth demand is *high* with up 1100 users. (a) Ratio shared revenue of WFP over ISP. (b) Average user utility. (c) The price that users pay for the service. (d) Amount of bandwidth sold to users.

**Third scenario**: we investigate the correlation between revenue sharing, pricing, users' utility, and bandwidth usage. The simulation setup includes an WFP providing 1000MB/sec to users with initial minimum price charged by an ISP of 10 units currency, minimum profit desired by the WFP is 5 units currency, users maximum willingness to pay is 100 units currency, and user utility and price are measured and determined using eq. (9) and (14) respectively. In this scenario, we consider two case studies: ($i$) When the WFP experiences *lower* and ($ii$) *higher* demand for bandwidth. **Case** ($\boldsymbol{i}$)**:** There are 10 to 100 users acquiring service from WFP. Figure 6(a) illustrates that shared revenue ratio of $\frac{WFP\ shared\ revenue}{ISP\ shared\ revenue}$ increases as the number of users increases, which confirms our previous simulation results. Figure 6(b) and Figure 6(c) demonstrate that users' average utility and price are stable when there is sufficient bandwidth for users. That is when the total bandwidth usage of 100 users described in Figure 6(d) is only 700 MB/sec $<$ 1000 MB/sec. **Case** ($\boldsymbol{ii}$): There are 100 to 1100 users subscribing from the WFP. The steep incline depicted in Figure 7(a) shows that the WFP rapidly achieves higher shared revenue as demands for bandwidth increase, and the WFP quickly attains near equal share as the ISP. In addition, the behavior of shared revenue illustrated in Figure 7(c) is also a reflection of price movement caused by the WFP hiking the price up when the total demands exceeds the capacity limit.

This leads to the bandwidth fluctuation illustrated in Figure 7(d) as a result from users adapting their demand for bandwidth when the price increases. In addition, Figure 7(a-c) reach the plateau (or flat) whenever bandwidth usage falls below the capacity limit, but change when demands go over the capacity limit. Moreover, Figure 7(c) also shows that user utility decreases as the price increases, because users obtain less bandwidth for higher price. Then, user utility in Figure 7(d) drops to zero when the price in Figure 7(b) peaks at 172 units currency, which results in unaffordable service leading to zero transaction. This also means no revenue for both the WFP and the ISP, as described in Figure 7(a). In conclusion, a WFP achieves higher shared revenue when there is high demand for bandwidth until the price becomes unaffordable, but this is also at the cost of lower user utility. On the other hand, we also demonstrate that a WFP also can obtain higher shared revenue and allow an ISP to gain higher revenue while achieving high user utility, when the bandwidth usage nears the limit capacity while keeping the price stable. Similar outcome is expected when an ISP price is increased beyond users' affordability except higher shared revenue will apportioned to an ISP relative to a WFP.

### B. *Revenue sharing between ISP and I-WFP*

Here, we investigate how the difference between the final sale price set by I-WFP to users and the minimum sale price set by ISP influences revenue sharing, and whether the outcome is favorable to ISP. Finally, the simulation is conducted to confirm that the scheme achieves the purposes mentioned in the previous section, i.e. to prevent I-WFP from achieving net profitability after deducting their monthly subscription fee. In order to clearly derive and depict the results, we have to avoid data noise and use the minimum sale price $g_{\bar{s}}^t$ and the final sale price $\lambda_{\bar{s}}^t$ after convergence.

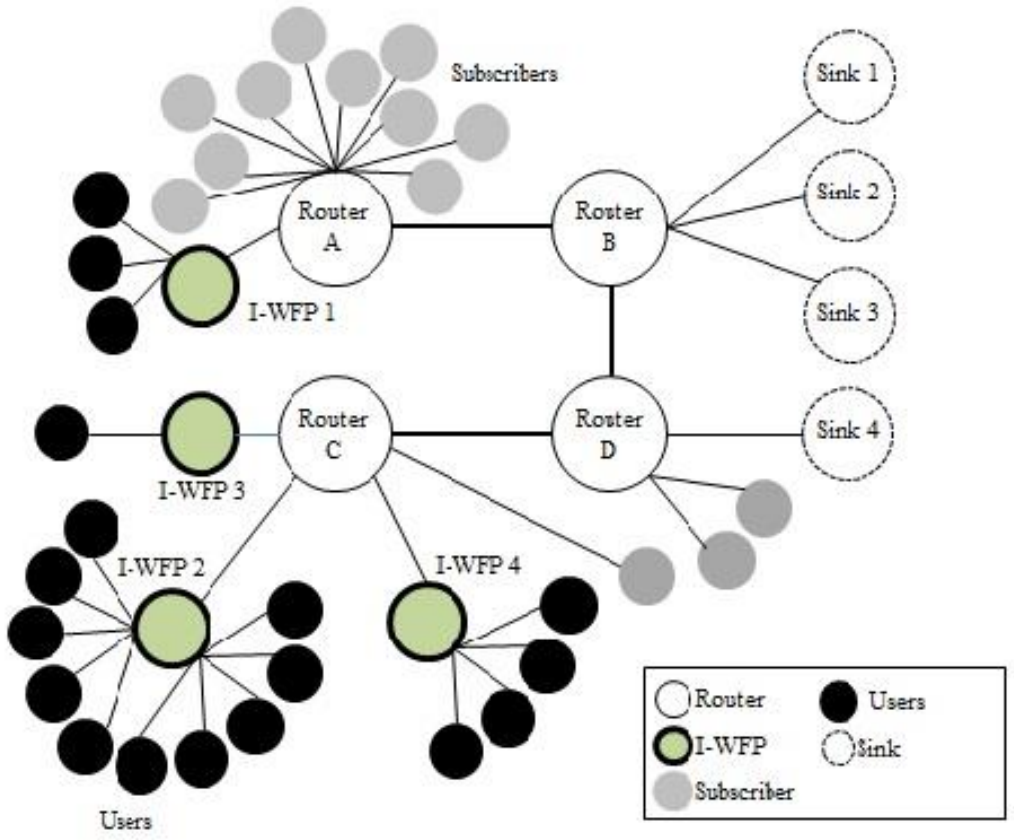

Fig. 6. Network Topology.

In this simulation, we employ a network with four nodes (node A, B, C, and D) connected by 3 links (link AB, CD, and BD) with a capacity of 50 MB each, as depicted in Figure 5. There are four I-WFPs providing connection service. Each I-WFP has a maximum of 200 MB in his/her data plan and he/she is hypothetically able to sell all of the bandwidth to users. In our setup, each I-WFP sells a maximum of 10 MB per transaction. I-WFP *one* is connected to node A and provides service to 3 users. There are 3 I-WFPs connected to node C: I-WFP *two* has an user, I-WFP *three* has 10 users, and I-WFP *four* has 4 users. There are 4 sinks (one, two, three, and four) that are connected to different users and subscribers. The path of each connection is described as follows:

- Users from I-WFP *one* is connected to *sink one* though link AB.
- Subscribers from router A are connected to *sink two* through link AB.
- Users of I-WFP *two* and *three* are connected to *sink four* through link CD.
- Users of I-WFP *four* are connected to *sink three* through link CD and DB.
- Subscriber from node C is connected to *sink four* through link CD.
- Subscribers from node D are connected to *sink three* through DB.

The simulation has three parts. First, ISP adjusts the minimum sale price according to the overall bandwidth demand by solving equation (5) and (6). Then, ISP presents the minimum sale price on each link to the I-WFPs in the network: link AB = 30 unit currency, CD = 3 unit currency, and BD = 1 unit currency. Link BD has the highest minimum price because there is a higher level of demand for bandwidth. Simultaneously and independently, I-WFPs compute their price, considering the bandwidth demand from interested users by solving equation (5) and (11). The four I-WFPs charge their users at 31, 26, 5, and 8 unit currencies respectively. Finally, ISP computes the revenue sharing between each I-WFP and ISP based on the specific revenue earned by each individual I-WFP.

| Prices | *I-WFP* | | | |
|---|---|---|---|---|
| | *1* | *2* | *3* | *4* |
| I-WFP's final sale price $\lambda$ (unit currency) | 31 | 26 | 5 | 8 |
| Minimum price to sell $g$ (unit currency) | 30 | 3 | 3 | 4 |

Table 1. Price set by I-WFPs and ISP.

The results of the revenue sharing between ISP and the 4 I-WFPs are depicted in Figures 3(a) – 3(d). Each unit on the x-axis represents 10 MB of bandwidth unused by I-WFP, and available for him/her to sell; the higher the number, the more bandwidth he/she still has to sell. For instance, when x = 12, there are still 120 MB available for sale and 80 MB already used. The y-axis is the percentage of revenue apportioned to the I-WFPs and ISP, totaling to 100%.

We observe several important outcomes. Firstly, from the graphs we can establish that the financial gain is higher for I-WFPs if they sell their bandwidth when they have more unused bandwidth than when they have less. This outcome confirms that the revenue sharing mechanism behaves as intended: as the amount of bandwidth used/sold increases, the revenue sharing mechanism discourages I-WFPs from selling their bandwidth by lowering their share of the revenue.

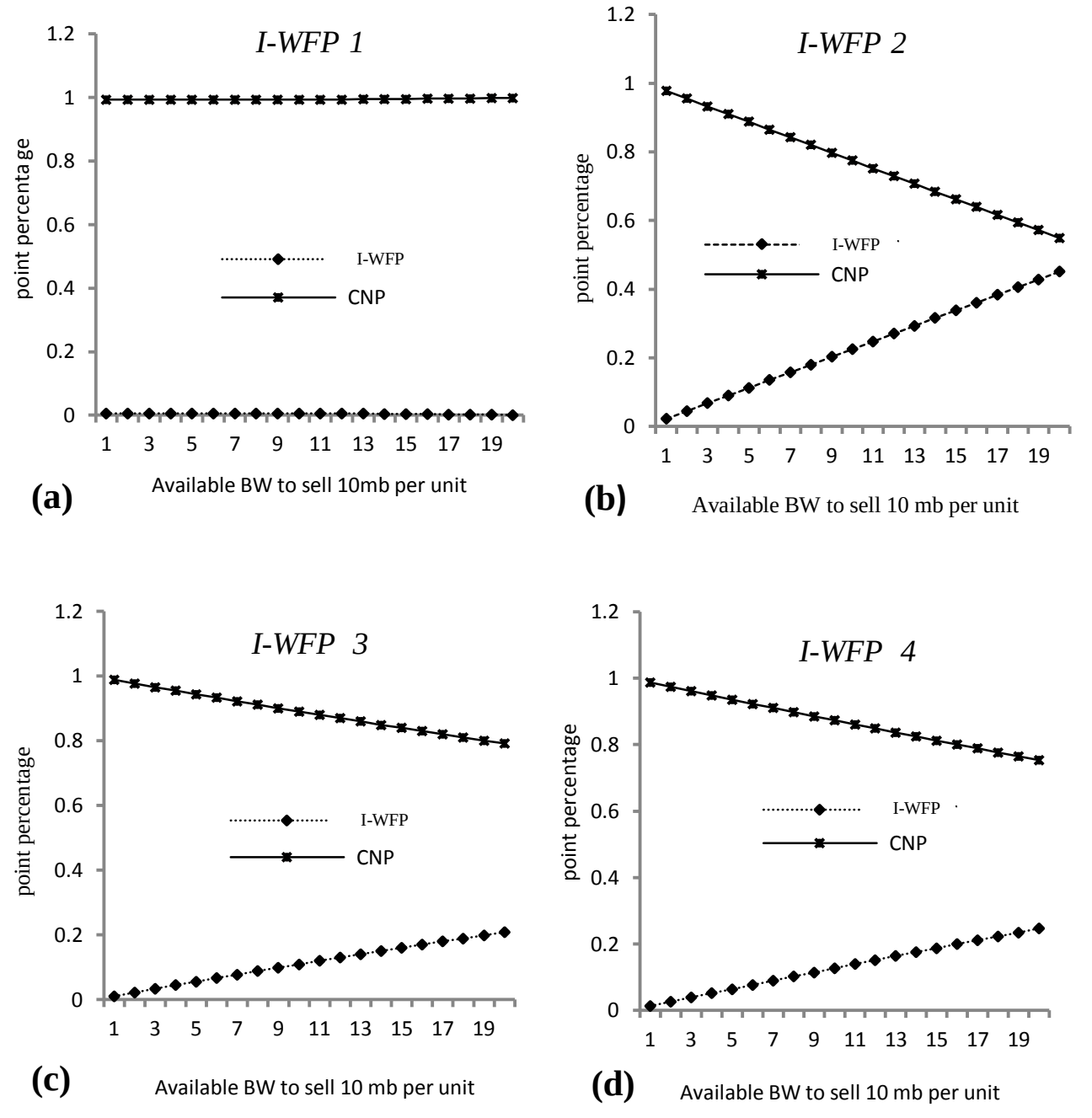

Fig. 7. Revenue sharing between each I-WFP and ISP according to Shapley value.

Secondly, the simulation outcome of I-WFP 1 (Figure 7(a)) proves that ISP can discourage bandwidth sale by increasing the minimum price when network experiences high traffic load during peak hours. In this scenario, the high minimum price (30 unit currency) is generated due to a high demand for bandwidth at link AB, including demand from non-selling subscribers. Since the request for bandwidth to I-WFP is low, I-WFP *one* is only able to set his/her price at 1 unit currency higher than the minimum sale price set by ISP (31 unit currency versus 30 unit currency).  This means I-WFP *one* contributes minimally to the transaction, hence receives only a negligible share of the revenue. This strategy also works to encourage I-WFPs to distribute the traffic load to non-peak hours to achieve better revenue share.

Different from I-WFP *one*, there are many users seeking to buy bandwidth from I-WFP *two*. This leads to the price set by I-WFP *two* to soar to 26 unit currency, while the minimum price for transmitting data on link CD is only set at 5 unit currency by ISP. The case of I-WFP *two* represents situations where ISP is experiencing low bandwidth demand but I-WFP is receiving a high level of demand. Hence, the minimum price set by ISP is much lower than the I-WFP's price to users. In such situations, the I-WFP has a higher bargaining power because of its higher "contribution" to the transaction. In recognition of this, ISP attributes a relatively higher share of the revenue gained from the transaction to the I-WFP. However, it is important to highlight here that even in such situations where I-WFP "contributes" significantly to the revenue gained, our revenue sharing scheme ensures that ISP still gains the majority share of the income, regardless of the amount of bandwidth a I-WFP has at that point of time, as illustrated. The next simulation confirms this effect in reverse: the percentage of the revenue share attributed to ISP increases as the difference between the minimum sale price and final sale price reduces, which characterize the cases with I-WFP *three* and *four* (Figure 7(c) and 7(d)).

This simulation confirms that the revenue sharing mechanism is favorable to ISP, as intended. It reduces I-WFP's portion of the revenue as more bandwidths are used/sold during each subscription cycle, even if the I-WFP is able to sell the bandwidth at a much higher price than the minimum price set by ISP. Sale of bandwidth is also discouraged by ISP during peak hours by raising the minimum sale price. The scheme also ensures ISP receives the majority portion of the revenue gained, regardless of the difference between the minimum sale price and the final sale price. At the same time, the scheme reasonably adjusts I-WFPs' portion of the revenue upward when the minimum price set by ISP is much lower than the I-WFP's final sale price. In the following simulation, we investigate the maximum share of revenue that a I-WFP may receive at different points of bandwidth usage within a subscription cycle. The simulation is conducted when the I-WFP's total bandwidth usage is at 0%, 25%, 50%, and 75%. In this simulation, we have a single link network connecting two nodes (node A and B) with a single I-WFP connected to node A. The I-WFP sells a total of 10 MB to *two* users. The total capacity of link AB is 20 MB per unit time. Next, we increase the I-WFP's price from 0 to 100 unit of currency when bandwidth usage is at 0, 25%, 50%, and 75%. Figure 8 illustrates the share of revenue received by the I-WFP approaches to near 50%, even when bandwidth usage is at 0%. However, I-WFP's portion of the revenue subsides as his/her bandwidth usage grows, as illustrated in figure 8. Conversely, the lowest portion of revenue ISP receives is at least 50% at all times.

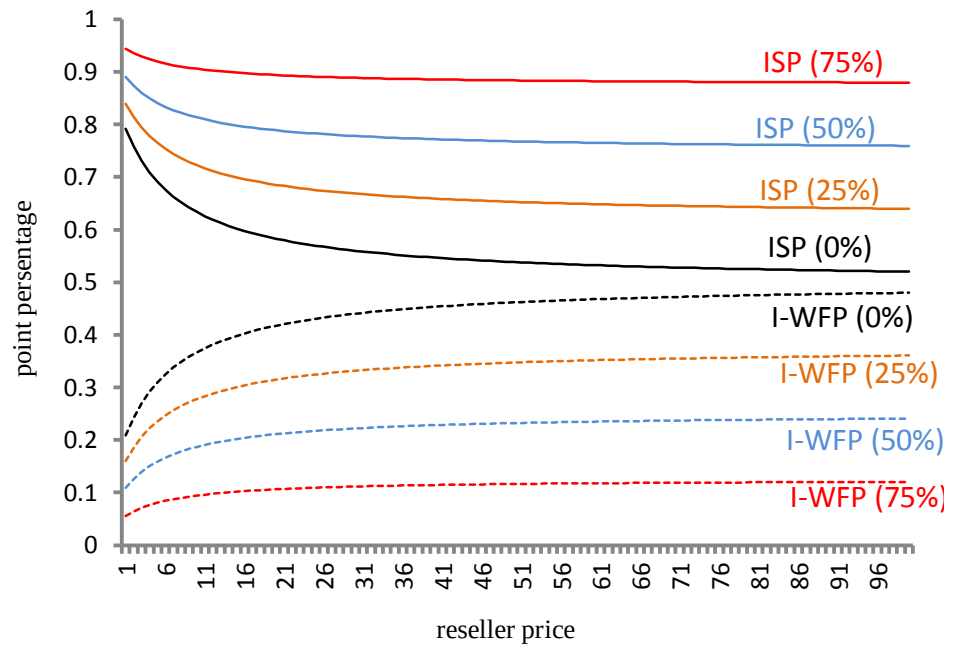

Fig. 8. Maximum and minimum shared received by I-WFPs and ISP.

The results from this simulation provide an insight as to when I-WFPs will most likely sell their bandwidths. Knowing this and given that ISP knows how many subscribers are at various stages of bandwidth usage in their subscription cycle, ISP is able to anticipate the additional traffic that may be generated from I-WFP's transactions, and devise better management so the connection quality of subscribers is not compromised. Over time, ISP may be able to estimate the size of such ad hoc demand by monitoring the total revenue generated by each I-WFP. Last but not least, it is also important to note that our model of bandwidth reselling and revenue sharing provides ISP with higher revenue from selling some of its bandwidth twice: first to subscribers, then to users through subscribers.

## VI. CONCLUSION

In this paper, we propose Shapley value based revenue sharing scheme and NUM based dynamic pricing strategy to leverage on the availability of E-WFP and I-WFP to provide Internet access to users. The revenue sharing and pricing mechanisms are designed to provide incentives for ISP, E-WFP and I-WFP to offer Additional bandwidth to their users. We demonstrate that, in cooperative revenue sharing, ISP will be aware of Wi-Fi providers' final sale price to users, and this gives ISP better understanding of the market and more control over pricing. Importantly, our revenue sharing model is able to address critical concerns such as traffic and congestion management. Specifically for E-WFP, our model also captures the conditions in which one of the parties (ISP or E-WFP) receives a higher portion of the revenue. For instance, the revenue sharing model apportions a larger share of the revenue to ISP when E-WFP generates lower revenue, but favors E-WFP when it contributes more to higher revenue. When E-WFP contributes a significant portion of the revenue, the division of revenue eventually converges to the region of 70%-30% division, the larger share given to E-WFP. Our dynamic pricing model and revenue sharing mechanism follow the economic concept of demand and supply, and critically considers each party's share of contribution. However, in all circumstances investigated, ISP is never disadvantaged in our revenue sharing model.

Contrary to the general perception that provision of on-demand network access service is disadvantageous to ISP, our pricing and revenue sharing model for I-WFP's sale of bandwidth to users demonstrates that it is possible for ISP to achieve financial gain. Importantly, by discouraging high data usage I-WFPs from selling and discouraging transactions during peak hours, our revenue sharing model is able to address critical concerns such as traffic management and the potential risk of ISP losing revenue. In addition, the scheme also prevents I-WFP from selling bandwidth for profit by adjusting the share of revenue apportioned to I-WFP according to the level of bandwidth usage and its "contribution" to the revenue. This model provides incentives for low data usage subscribers to keep their subscription by providing them with an opportunity to earn some income to offset their subscription fee.

In our future work, we will investigate whether the economic interplay and negotiation between ISP, Wi-Fi providers, and users reach an equilibrium. If it does, in what condition the equilibrium is reached and how it impacts the revenue sharing mechanism. In addition, in order to design pricing and revenue scheme that reflect quality of service, the role QoE in the design will be addressed as well.